%Paper: gr-qc/9405055
%From: hayward@murasaki.scphys.kyoto-u.ac.jp
%Date: Wed, 25 May 94 15:00:04 JST

\font\lbf=cmbx10 scaled\magstep2

\def\bs{\bigskip}
\def\ms{\medskip}
\def\np{\vfill\eject}

\def\ni{\noindent}
\def\cl{\centerline}

\def\ref#1#2#3#4{#1\ {\it#2\ }{\bf#3\ }#4\par}
\def\refb#1#2#3{#1\ {\it#2\ }#3\par}

\def\CPAM{Comm.\ Pure App.\ Math.}
\def\PR{Phys.\ Rev.}
\def\PRL{Phys.\ Rev.\ Lett.}

\def\O#1{\left.#1\right\vert_S}
\def\H#1{\left.#1\right\vert_H}
\def\I{{\cal I}}
\def\L{{\cal L}}

\magnification=\magstep1

\cl{\lbf Confinement by black holes}
\ms\cl{\bf Sean A. Hayward}
\ms\cl{Department of Physics}
\cl{Kyoto University}
\cl{Kyoto 606-01}
\cl{Japan}
\ms\cl{24th May 1994}
\ms\ni
{\bf Abstract.}
The question of whether an observer can escape from a black hole is addressed,
using a recent general definition of a black hole in the form of
a future outer trapping horizon.
An observer on a future outer trapping horizon
must enter the neighbouring trapped region.
It is possible for the observer to subsequently escape from the trapped region.
However,
if the horizon separates the space-time into two disjoint components,
inside and outside the horizon,
then an observer inside a future outer trapping horizon cannot get outside,
assuming the null energy condition.
A similar confinement property holds
for trapped, locally area-preserving cylinders,
as suggested by Israel.
\bs\ni
It is widely believed that anyone falling into a black hole
is forever confined within the black hole
and cannot escape back to the universe outside.
A precise general formulation of this conjecture
has been impossible until recently,
due to the lack of a general definition of a black hole.
The conventional definition, the event horizon,
is defined if the space-time is asymptotically flat,
but the actual universe is thought not to be asymptotically flat.
A definition of black hole in arbitrary space-times has recently been given
in the form of a {\it future outer trapping horizon} [1].
This article addresses the question of whether
an observer inside such a horizon can get outside.
This could be stated more geometrically
in terms of future-directed temporal curves,
along which observers move.
\ms
Recall the definition of a black hole [1].
The space-time is assumed time-orientable.
Given a double-null foliation of spatial 2-surfaces with area form $\mu$,
the expansions $\theta_\pm$ may be defined by $\mu\theta_\pm=\L_\pm\mu$,
where $\L_\pm$ denotes the Lie derivative
in the future-pointing null normal directions.
A {\it marginal surface} is a spatial 2-surface $S$
on which one expansion vanishes, fixed henceforth as $\O{\theta_+}=0$.
A {\it trapping horizon} is the closure $\overline{H}$ of a 3-surface $H$
foliated by marginal surfaces on which
$\H{\theta_-}\not=0$ and $\H{\L_-\theta_+}\not=0$,
where the double-null foliation is adapted to the marginal surfaces.
The trapping horizon is said to be
{\it outer} if $\H{\L_-\theta_+}<0$,
{\it inner} if $\H{\L_-\theta_+}>0$,
{\it future} if $\H{\theta_-}<0$ and
{\it past} if $\H{\theta_-}>0$.
For a future outer trapping horizon,
the idea is that the outgoing light rays
are diverging just outside the horizon and converging just inside,
and that the ingoing light rays are converging.
This provides the definition of a black hole.
Similarly, a past outer trapping horizon
provides the definition of a white hole.
Inner trapping horizons include cosmological horizons
as well as the possible inner boundaries of black or white holes.
\ms
Recall also that a {\it trapped surface} [2] is
a compact spatial 2-surface $S$ such that $\O{\theta_+\theta_-}>0$,
and is said to be {\it future trapped} if $\O{\theta_\pm}<0$
and {\it past trapped} if $\O{\theta_\pm}>0$ [1].
A {\it trapped region} [1] is a connected subset of space-time
for which each point lies on some trapped surface.
A basic property of a trapping horizon is that
if the foliating marginal surfaces $S$ are compact,
any spatial 2-surface sufficiently close to $S$
is trapped if it lies to one side of the horizon,
and is not trapped if it lies to the other side [1].
In other words,
there is a neighbouring trapped region
whose boundary includes the trapping horizon.
\ms\ni
{\it Proposition 1.}
Assuming the null energy condition,
an observer on a future outer trapping horizon
must enter the neighbouring trapped region.
Similarly, an observer on a past outer trapping horizon
must have come from the neighbouring trapped region.
\ms\ni
{\it Proof.}
The signature law [1] states that an outer trapping horizon is spatial or null,
assuming the null energy (or convergence) condition [2--3],
which is implied by the weak energy condition.
Thus an observer on the horizon must cross it.
For a future or past, outer trapping horizon,
the neighbouring trapped region is to the future or past respectively [1].
\ms
This shows that outer trapping horizons act as one-way membranes,
across which an observer may move in one direction only.
This does not necessarily mean that
an observer cannot get back to the other side,
since there may be an alternative route which does not cross the horizon,
which may be thought of as a type of wormhole.
An example is the Reissner-Nordstr\"om black hole,
periodically identified in time, as in Figure 1(a).
In this case,
an observer may enter and leave the trapped region arbitrarily often.
Any confinement theorem for black holes must exclude such examples.
One might wish to exclude the above example on the grounds that
it contains causal loops,
but this would be an unnecessarily strong restriction on allowable space-times.
Alternatively,
note that the problem with this example is that the horizon is not achronal.
Recall that an {\it achronal} (or semispacelike) hypersurface [2--3]
is one for which no two points can be connected by a temporal curve,
or more prosaically,
a hypersurface which an observer cannot cross more than once.
\ms
In order to formulate a statement about confinement,
one would like to be able to distinguish the interior and exterior
of the black hole.
Consider the complement of a hypersurface in the space-time,
which consists of either one or two disjoint components.
If there are two components,
the hypersurface will be called {\it separating}.
(Note that hypersurfaces with boundary are not separating).
If the hypersurface is not separating,
it does not make sense to say that
the complement is either inside or outside the hypersurface.
In the above example, the trapping horizon is not separating,
and there is no `interior' to which observers might be confined.
For a separating trapping horizon,
the component containing the neighbouring trapped surfaces
will be called the {\it interior}
and the other component the {\it exterior} of the horizon.
Similarly,
points in the interior will be described as {\it inside} the horizon,
and points in the exterior as {\it outside} the horizon.
\ms\ni
{\it Confinement theorem I.}
Assuming the null energy condition:
(i) an observer inside a separating future outer trapping horizon
cannot get outside;
(ii) an observer outside a separating past outer trapping horizon
cannot get inside;
(iii) a separating outer trapping horizon is achronal.
\ms\ni
{\it Proof.}
For a separating horizon,
any path between inside and outside must cross the horizon,
and according to Proposition 1,
an observer may cross a future outer trapping horizon
from outside to inside only.
Thus an observer inside the horizon cannot get outside,
and the horizon is achronal.
Similarly for a past outer trapping horizon.
\ms
Mathematically, the confinement theorem is a simple consequence
of the properties of trapping horizons derived in [1],
which illustrates the usefulness of the definition of trapping horizon.
As an example, consider a typical asymptotically flat,
spherically symmetric black hole according to cosmic censorship,
as depicted in Figure 1(b).
This picture has been verified
for generic configurations of a massless scalar field by Christodoulou [4--5].
The basic elements in the picture are:
conformal infinity $\I^+$, which is null;
the regular centre $C_R$, which is temporal;
the central singularity $C_S$, which is spatial or null;
and the future outer trapping horizon $\overline{H}$,
or apparent horizon,\footnote\ddag
{Recall that a general space-time does not have a unique apparent horizon,
since the definition [3] is foliation-dependent [6].
However, in spherical symmetry,
an asymptotically flat space-time
has a uniquely defined maximal spherically symmetric apparent horizon
for each component of $\I^\pm$,
which together coincide with the outermost trapping horizon(s).}
which is spatial or null.
Also drawn are the 3-cylinders of constant area,
which are spatial inside the trapped region and Lorentzian outside,
as explained below.
The area of the trapping horizon is non-decreasing [1].
So in this example,
the trapping horizon is separating,
and its interior is exactly the trapped region.
According to the cosmic censorship conjecture,
such a scenario is typical of generic non-symmetric gravitational collapse.
\ms
Incidentally, the reason for defining the trapping horizon as a closure
is to close up gaps left by degenerate marginal surfaces or isolated points,
as in Figure 1(a) and (b) respectively.
Note also that while the trapping horizon in Figure 1(a)
is neither separating nor achronal,
the corresponding horizon
in the maximally extended Reissner-Nordstr\"om space-time
is separating and achronal.
In the latter case,
an observer inside the horizon cannot get outside,
but may still leave the trapped region.
This illustrates that one may also be interested in
confinement to a particular trapped region.
This question will not be addressed here,
except to note that escaping from the trapped region requires
the formation of another trapping horizon inside the black hole.
\ms
The confinement theorem requires a global assumption on the space-time,
namely that the trapping horizon is separating.
There are no other global assumptions,
and in particular no causality assumptions.
For instance,
there could be a region of causal loops to the future of the horizon.
Nevertheless,
one might wonder whether one of the standard causality conditions
could replace that of a separating trapping horizon.
Global hyperbolicity or the stable causality condition [3] would suffice,
for then any spatial or null hypersurface is achronal.
However, this would be a rather strong restriction.
On the other hand, the chronology condition, causality condition,
future or past distinguishing condition and strong causality condition [3]
all allow non-achronal spatial hypersurfaces,
as in examples similar to Figure 38 of [3].
Since an observer could cross such a trapping horizon twice,
there is no real distinction between inside and outside.
So it seems that these causality conditions
are not appropriate to the problem of confinement.
\ms
Israel suggested a different construction
to investigate confinement by black holes [7].
This involves a generalization of
the constant-area cylinders of a spherically symmetric space-time,
namely a {\it locally area-preserving cylinder},
defined as a hypersurface foliated by spherical spatial 2-surfaces
whose area form $\mu$ is preserved in the direction $v$
normal to the 2-surfaces and tangent to the hypersurface.
If the foliating 2-surfaces are (future or past) trapped surfaces,
the cylinder will also be described as (future or past) trapped.
\ms\ni
{\it Proposition 2.}
A locally area-preserving cylinder is spatial if and only if it is trapped.
\ms\ni
{\it Proof.}
The direction $v$
is a linear combination of the future-pointing null normals $l_\pm$,
$v=\beta l_+-\alpha l_-$,
so that $0=\L_v\mu=\mu(\beta\theta_+-\alpha\theta_-)$.
The condition defining a trapped surface, $\theta_+\theta_->0$,
is then equivalent to the condition that $v$ is spatial, $\alpha\beta>0$.
\ms
This allows a similar confinement theorem for such cylinders.
With respect to a separating, future trapped, locally area-preserving cylinder,
points to the future or past may be described as
inside or outside respectively;
similarly for a past trapped cylinder.
Then the same argument as in the proof of the confinement theorem
yields the following.
\ms\ni
{\it Confinement theorem II.}
(i) An observer inside
a separating, future trapped, locally area-preserving cylinder
cannot get outside;
(ii) an observer outside
a separating, past trapped, locally area-preserving cylinder
cannot get inside;
(iii) a separating, trapped, locally area-preserving cylinder is achronal.
\ms
However, this is not so useful,
since locally area-preserving cylinders are unlikely to be
both trapped and separating in a typical gravitational collapse.
For instance, in the spherically symmetric collapse of Figure 1(b),
the constant-area cylinders are all partially outside the trapped region,
where they are Lorentzian rather than spatial,
by the argument of Proposition 2.
One could patch this up by taking the trapped part of the cylinder
and extending it along any spatial hypersurface outside the trapped region
which includes the centre,
or by considering the `interior' to be
the causal future of the trapped part of the cylinder,
or by some other such modification.
This will not be pursued explicitly here because
the confinement theorem for trapping horizons already meets the need.
\ms
In conclusion, it has been shown that a black hole by the new definition
possesses the inescapability
which is often said to characterize black holes,
given a global assumption on the space-time to the effect that
the interior and exterior of the trapping horizon are distinct.
This assumption excludes certain sorts of wormholes,
the existence of which is still open to question.
So this has not resolved the question of
the fate of observers falling into a black hole,
but rather has clarified the assumptions necessary to ensure confinement.
The assumptions have been found to be quite weak,
namely the null energy condition
and the disjointness of the black hole's interior and exterior.
Finally, it should be emphasized that the results are valid in any space-time,
not necessarily asymptotically flat, nor free of naked singularities,
nor free of causal loops.
In particular, the confinement theorem is valid in cosmological space-times,
and so applies to actual astrophysical black holes.
\bs\ni
Acknowledgements.
It is a pleasure to thank
Marcus Kriele, Ken-ichi Nakao and Tetsuya Shiromizu for discussions,
and the Japan Society for the Promotion of Science for financial support.
\np
\begingroup
\parindent=0pt\everypar={\hangindent=20pt\hangafter=1}\par
{\bf References}\ms
\ref{[1] Hayward S A 1994}{General laws of black-hole dynamics, \PR}{D49}
{(in print)}
\refb{[2] Penrose R 1968 in}{Battelle Rencontres}
{ed: DeWitt C M \& Wheeler J A (Benjamin)}
\refb{[3] Hawking S W \& Ellis G F R 1973}
{The Large Scale Structure of Space-Time}{(Cambridge University Press)}
\ref{[4] Christodoulou D 1991}\CPAM{44}{339}
\ref{[5] Christodoulou D 1993}\CPAM{46}{1131}
\ref{[6] Wald R M \& Iyer V 1991}\PR{D44}{R3719}
\ref{[7] Israel W 1986}\PRL{56}{789}
\endgroup
\bs\ni
{\bf Figure 1.}
Conformal diagrams of:
(a) the Reissner-Nordstr\"om space-time,
with topological identifications along the notched hypersurfaces;
(b) a typical asymptotically flat,
spherically symmetric gravitational collapse with regular centre $C_R$,
according to cosmic censorship.
The central singularities are labelled $C_S$,
the heavy line represents the future outer trapping horizon $H$,
and the lighter curves represent the constant-area cylinders.
\bye